\documentclass[10pt]{article}
\usepackage{latexsym}
\usepackage[dvips]{graphicx}
\usepackage{overcite}
\usepackage{amssymb}
\usepackage{amsbsy}
\usepackage{amsmath}
\usepackage{mathrsfs}
\usepackage{xr}
\usepackage{booktabs} 
\usepackage{multirow} 
\usepackage{xcolor}	  
\usepackage{mathtools}
\usepackage{graphicx}
\usepackage{tabularx}
\usepackage[labelformat=simple, font=small,labelfont=bf]{caption}
\usepackage{placeins}
\usepackage{authblk}

\graphicspath{ {./plot/} }

\DeclareRobustCommand*{\Figure}[3]{
   \begin{figure}[!htb]
   \begin{center}
   \noindent
   \includegraphics[width=#2]{#1}  
   \end{center}
   \caption{#3}
   \addtocontents{lof}{\vspace{\baselineskip}}
   \label{fig:#1}
   \end{figure}
}

\newcommand{\be}{\begin{equation}}
\newcommand{\ee}{\end{equation}}

\begin{document}

\title{Micelle Formation of Diblock Copolymers Driven by Cononsolvency Effect}

\author[1]{Xiangyu Zhang\footnote{E-mail: xzhan357@jh.edu}}
\author[2]{Jing Zong}
\author[3]{Dong Meng}
\affil[1]{Department of Chemical and Biomolecular Engineering, John Hopkins University, Baltimore, MD 21218, United States}
\affil[2] {Dave C. Swalm School of Chemical Engineering, Mississippi State University, MS 39762, United States}
\affil[3]{Biomaterials Division, Department of Molecular Pathobiology, New York University, New York, NY 10010, United States}

\maketitle

\begin{abstract}
This study utilizes self-consistent field theory to characterize various features of cononsolvency-driven spherical micelles formed by double hydrophilic block copolymers (DHBCs). Micelles are observed only at an intermediate cosolvent fraction, forming abruptly at a specific solvent/cosolvent mixing ratio and gradually disappearing with further cosolvent addition. A stronger core-block – cosolvent attractive interaction leads to a lower critical micelle concentration and a higher aggregation number.
The density profile of cononsolvency-driven micelles is compared with that of conventional micelles, which form due to core-block – solvent repulsive interactions. In conventional micelles, the core is primarily occupied by polymer segments, whereas in cononsolvency-driven micelles, the core consists mainly of solvents and cosolvents. This fundamental difference can be explained through thermodynamic analysis. Conventional micelle formation is driven by the reduction of core-block – solvent contact due to repulsive interactions. In contrast, cononsolvency-driven micelle formation is governed by an increase in core-block – cosolvent contact area, playing the major role to minimize the total free energy--an essential distinction from conventional micelles.

\end{abstract}

\section{Introduction}
The ability of diblock copolymers to self-assemble into ordered-structure has caught great interest due to its broad applications in drug delivery, nanoreactors, catalysis, etc. \cite{sistach2010thermoresponsive, khimani2020self,wei2009thermo,ge2007stimuli,li2008self,chung1999thermo,zhang2014polymer,schmidt2018double} The common environmental factors to stimulate micelle structure changes are temperature, pH, sound, light and solvent compositions.\cite{papadakis2019switch,qian2019new,liu2014dual,guragain2015multi,rapoport2007physical,wang2008investigation} However, the use of solvent composition may give rise to some "counter-intuitive" behaviors in both homopolymer system and diblock copolymer system, for which our understanding is still lack. \cite{wolf1975measured,wolf1978measured,rao2007cononsolvency,michailova2010nanoparticles,kyriakos2016quantifying,ko2021co}
\par
For polymer immersed in binary solvent mixtures, two intriguing phenomena are often discussed, cosolvency and cononsolvency. Cosolvency means that the mixture of two poor solvents can enhance the polymer solubility. \cite{dudowicz2015communication} The common example system is poly(methyl methacrylate)(PMMA) immersed in water+alcohol mixture, which is a UCST type polymer. \cite{cowie1987alcohol} Water and alcohol are both non-solvents for PMMA. But the mixture of them can enhance the solubility at intermediate composition range, corresponding to the decrease of the critical temperature where phase separation can occur.\cite{cowie1987alcohol} Cononsolvency means that the mixture of two good solvents can create a bad solution condition for the polymer, corresponding to the decrease of the miscibility. \cite{dudowicz2015communication} The common example system is PNIPAm immersed in water+alcohol mixture, which is a LCST type polymer. The significant collapse of the PNIPAm chain can be observed in the water/alcohol mixture, whereas the chain exhibits extended state in pure water or pure alcohol. \cite{wang2012conformational, zhang2001reentrant} So, the hydrophobic interaction to induce micro-structure may be substituted by cononsolvency effect through tuning solvent compositions. 
DHBCs have several advantages over common amphiphilic block copolymers, such as biocompatibility, permeability, degradability.\cite{knop2010poly,schmidt2018double} It has been reported that the ordered structure formed by DHBCs can be observed in neat solvents without the external stimulation.\cite{casse2012solution,ke2009association,rudolph2013synthesis,willersinn2017aqueous,park2016micelles,willersinn2017self} However, the low self-assembly efficiency of DHBCs and the instability of the resulting microstructures present major challenges for their potential applications.\cite{schmidt2018double} Thus, utilizing a binary solvent mixture to induce the cononsolvency effect may be a feasible approach to overcoming the aforementioned limitations, as the strength of the cononsolvency effect can be tuned by adjusting the cosolvent fraction. 
\par
In general, cononsolvency effect on the micelle formed by diblock copolymers can be categorized into two types. One is micelle formation driven by cononsolvency, in which micelle is composed of DHBCs.\cite{rao2007cononsolvency,ge2007stimuli,michailova2010nanoparticles,michailova2018self,wu2022synthesis} For example, poly(N-isopropylacrylamide)-b-poly(oligo(ethylene glycol) methyl ether methacrylate)(PNIPAM-b-POEGMA) diblock copolymers are unimers in pure water or pure methanol below LCST. But PNIPAm-core micelle can be formed in methanol/water mixture.\cite{rao2007cononsolvency}  The other type is micelle morphology modified by cononsolvency, in which micelle is composed of one permanent hydrophobic block and one PNIPAm block, and PNIPAm constitutes as the micelle shell. \cite{kyriakos2014cononsolvency,wang2008investigation,kyriakos2016quantifying,zhou2016pfs,ko2021co} For example, the PNIPAm shell of the micelle formed by Poly(methyl methacrylate)‑b‑poly(N‑isopropylacrylamide)(PMMA-b-PNIPAM) will become shrank with the addition of the methanol, and the methanol fraction should not be too high as cosolvency of PMMA will come into effect, which will cause the dissolve of the polymer.\cite{ko2021co} In this study, the self-assembly of DHBCs induced by solvent compositions is investigated. 
\par
Accordingly, the mechanism of cononsolvency should first be discussed, though it is still under debate. Bharadwaj \textit{et al.} categorized current proposed driving force for cononsolvency into four aspects, (a) cosolvent-solvent attraction, (b) enthalpic bridging, (c) geometric frustration, (d) cosolvent surfactant mechanism.\cite{bharadwaj2022cononsolvency} Full details can be found in that paper, which will not be elaborated here. The important point is that (b)(c)(d) actually can be generalized as one effect, which is the strength of polymer-cosolvent affinity force, no matter whether it is driven by entropy or enthalpy.\cite{zhang2020unified, zhang2024general} So, the mechanism of cononsolvency becomes the question whether cosolvent prefer more contacting with polymer or more with solvents, corresponding to P-C driven and S-C driven system, respectively. \cite{zhang2020unified, zhang2024general} For cononsolvency driven micelle, both blocks can be the core if cononsolvency effect was driven by S-C attraction, but that is not the case in experiments. Thus, the model to study cononsolvency effect on micelle behaviors should be built up based on P-C attractive interaction. 
\par
The paper is organized as following. First, the boundary of homogeneous phase to micellar phase transition induced by cononsolvency is calculated by self-consistent field theory, which is the binodal boundary. And the aggregation number of B-block segments at critical point is shown. Next, the density profile of the micelle induced by cononsolvency is analyzed, and it is compared with that of the micelle driven by solvent selectivity. At last, the driving force for the micellization is discussed in both conventional micelle and cononsolvency micelle system.
\par
\section{Model and Method}
We consider a system containing solvents(S), cosolvents(C) and diblock copolymer chains(A-b-B) with length of each block being $N_{A}=32$ and $N_{B}=16$ at temperature $T$ in constant volume $V$ and constant chemical potential $\mu_{P}$. The chain length of the polymer is $N_{P}=N_{A}+N_{B}$. Non-bonded potential is described by Flory-Huggins $\chi$, and bonded potential is given by discrete gaussian bond. \par
The partition function in grand canonical ensemble of A-B/S/C system can be written in the form,
\begin{equation}
\begin{split}
{\Xi}(\mu_{P},\mu_{S},\mu_{C},V,T) = & \sum_{n_{P}=0}^{\infty} \sum_{n_{S}=0}^{\infty} \sum_{n_{C}=0}^{\infty} \lambda_{T}^{-3n_{P}N-3n_{S}-3n_{C}} e^{\mu_{P}n_{P}+\mu_{S}n_{S}+\mu_{C}n_{C}} \frac{1}{(n_{\rm P})! n_{\rm S}! n_{\rm C}!} \\
& \prod_{j=1}^{n_{\rm S}}\int {\rm{d}} {{\bf r}_{{\rm S},j}} \prod_{j'=1}^{n_{\rm C}}  \int {\rm{d}} {{\bf r}_{{\rm C},j'}}\prod_{k=1}^{n_{\rm P}}  \prod_{s=1}^{N_P}  \int {\rm{d}} {{\bf R}_{k,s}}  \exp \left( - \beta \mathcal{H}^{\rm{b}} - \beta \mathcal{H}^{\rm{nb}} \right)
\end{split}
\end{equation}
where the Hamiltonian due to the bonding interaction is given by,
\be
\mathcal{H}^b=\sum_{k=1}^{n_{P}}\sum_{s=1}^{N_{P}-1} \frac{3k_BT}{2a^2} \left| {\bf R}_{i,s} - {\bf R}_{i,s+1} \right|^2
\ee
And the Hamiltonian due to the non-bonded interaction is given by,
\be
\mathcal{H}^{\rm{nb}} = \frac{1}{2} \sum\limits_{\alpha = {\rm{P, S, C}}} \sum\limits_{\alpha' \neq \alpha} \int{\rm d}{\bf r}\int{\rm d}{\bf r}' \hat{\phi}_{\alpha} ({\bf r})  u_{\alpha \alpha '}({\bf r}, {\bf r}') \hat{\phi}_{\alpha '} ({\bf r}')
\ee
with $u_{\alpha \alpha '}({\bf r}, {\bf r}')=\chi_{\alpha\alpha'}\delta({\bf r}-{\bf r'})$ and the microscopic number densities of P and S(C) segments at spatial position ${\bf r}$ defined as
\be
\hat{\phi}_{\rm P} ({\bf r}) \equiv \sum_{k=1}^{n_{\rm P}} \sum_{s=1}^{N_P} \delta ({\bf r} - {\bf R}_{P,(k,s)}),
\ee
\be
\hat{\phi}_{\rm S (C)} ({\bf r}) \equiv \sum_{s=1}^{n_{\rm S (C)}} \delta ({\bf r} - {\bf r}_{{\rm S (C)},s}),
\ee
By inserting the identity
\[
1 = \prod_{\alpha={\rm A,B,S,C}} \int \mathscr{D}\phi_{\alpha} \mathscr{D}\omega_{\alpha} \exp \left\{ \int \mathrm{d} \mathbf{r} \omega_{\alpha}(\mathbf{r}) \left[ \phi_{\alpha}(\mathbf{r}) - \hat{\phi}_{\alpha}(\mathbf{r}) \right] \right\},
\]
, where $\omega_{\alpha}(\mathbf{r})$ is the purely imaginary conjugate field interacting with species $\alpha $, and applying the saddle point approximation, the SCFT equation is given as following,
\begin{equation}
\omega_{\alpha}({\bf r})=\sum_{\alpha'}^{\alpha\neq\alpha'}\chi_{\alpha\alpha'}\phi_{\alpha'}({\bf r})+\xi({\bf r})
\end{equation}
\begin{equation}
\phi_{S}({\bf r})=z_{S}\exp(-\omega_{S}({\bf r}))
\end{equation}
\begin{equation}
\phi_{C}({\bf r})=z_{C}\exp(-\omega_{C}({\bf r}))
\end{equation}
\begin{equation}
\phi_{A}({\bf r})=z_{P}\exp(\omega_{A}({\bf r}))\sum_{s=1}^{N_{A}}q_{s}({\bf r})q_{s}^{*}({\bf r})
\end{equation}
\begin{equation}
\phi_{B}({\bf r})=z_{P}\exp(\omega_{B}({\bf r}))\sum_{s=1}^{N_{B}}q_{s}({\bf r})q_{s}^{*}({\bf r})
\end{equation}
\begin{equation}
\begin{split}
\xi({\bf r}) = & \omega_{C}({\bf r})-\chi_{BC}(1-\phi_{A}({\bf r})-\phi_{S}({\bf r})-\phi_{C}({\bf r}))-\chi_{AC}(1-\phi_{B}({\bf r})-\phi_{S}({\bf r})-\phi_{C}({\bf r})) \\ 
& -\chi_{SC}(1-\phi_{A}({\bf r})-\phi_{B}({\bf r})-\phi_{C}({\bf r}))
\end{split}
\end{equation}
where $q({\bf r}, s)=\exp(-\omega_{\alpha}({\bf r})) \int {\rm d} {\bf r'} \Phi(\left | {\bf r}-{\bf r'}\right |)q({\bf r'}, s-1)$, $s \leq N_{A}, \alpha=A; s > N_{A}, \alpha=B$ and $q^{*}({\bf r}, N_{P}-s+1)=\exp(-\omega_{\alpha}({\bf r})) \int {\rm d} {\bf r'} \Phi(\left | {\bf r}-{\bf r'}\right |)q({\bf r'}, N_{P}-s+2)$, $s \leq N_{A}, \alpha=B; s > N_{A}, \alpha=A$, are the chain propagators starting from the first and the last segments, respectively. And $\Phi$ is the bond transition factor, $\Phi(\left | {\bf r}-{\bf r'} \right |)=(\frac{3}{2 \pi a^2})^{\frac{3}{2}} \exp(-\frac{3r^2}{2a^2})$. $Q_P$ is the single chain partition function, $Q_P=1/V \int dr \exp(\omega_{A}(r))q(r,1)q^{*}(r,N_{P})$. 

$z_{\alpha}$ is the activity of $\alpha$ component, which is coupled to chemical potential, $\phi_{\alpha}(r)$ is the $\alpha$ component volume fraction at $r$ position, $\chi_{\alpha\alpha'}$ describes the interaction strength between different species, if $\alpha=\alpha'$, $\chi_{\alpha\alpha'}=0$, $\xi$ is the the external potential to ensure the incompressibility condition. Different from common treatment that $\xi$ expression derived by algebra manipulation, we substitute $\phi_{A}(r)+\phi_{B}(r)+\phi_{S}(r)+\phi_{C}(r)=1$ condition into $\omega_{C}(r)$ equation to obtain $\xi$. The reason is that some of $\chi_{\alpha \alpha'}$ being $0$ leads to the incapability to find $\xi$ solution. \par 
Next the system is reduced to one dimension in spherical coordinates by assuming $\psi$ and $\theta$ are constants. The integration of propagator in one dimension can be written as,
\begin{eqnarray}
q(r,s) 
& = & \exp(-\omega_{P}(r)) \int_{0}^{L_{r}}{\rm d}{r'} \int_{0}^{\pi}{\rm d}{\theta'} \int_{0}^{2 \pi}{\rm d}{\psi'} \sin(\theta') {{r'}^2} (\frac{3}{2 \pi a^2})^{\frac{3}{2}} \nonumber \\ & & \exp(-\frac{3}{2 a^2} (r^2 + {r'}^2 - 2 rr' \cos(\theta')) q(r', s) \nonumber \\
& = & (\frac{3}{2\pi a^2})^{\frac{1}{2}} \exp(-\omega_{P}(r)) \int_{0}^{L_{r}} {\rm d}{r'} \frac{r'}{r} (\exp(-\frac{3(r-r')^2}{2a^2}) - \exp(-\frac{3(r+r')^2}{2a^2})) \nonumber \\ & & q(r', s-1)
\end{eqnarray}
Finally, the free energy of the system is,
\begin{eqnarray}
& & \mathcal{H}^{G} [\phi_{A},\phi_{B},\phi_{S},\phi_{C},\omega_{A},\omega_{B},\omega_{S},\omega_{C}]= \nonumber \\
& & \frac{1}{2} \sum_{\alpha =A,B,S,C} \sum_{\alpha' =A,B,S,C} \int d{\bf r} \int d{\bf r'}\phi_{\alpha}({\bf r})u_{\alpha \alpha'}({\bf r}-{\bf r'})\phi_{\alpha'}({\bf r'})- \sum_{\alpha= A,B,S,C}\int d{\bf r} \phi_{\alpha}({\bf r})\omega_{\alpha}({\bf r})  \nonumber \\
& & 
-  z_{P} V Q_{P}[\omega_{A},\omega_{B}] -  z_{S}V Q_{S}[\omega_{S}] - z_{C}V Q_{C}[\omega_{C}]
\end{eqnarray}
\par
To make sure that micelle formation is purely driven by cononsolvency effect, all other parameters, except for $\chi_{BC}$, are set to $0$. Because of the block length and $\chi$ values we choose, B-block will aggregate as the micelle core.
The critical point is defined as the $\phi_{P}^{cr}$ where grand potential of the inhomogeneous system equals to the grand potential of homogeneous system(constant solution). The interface of the micelle is decided by the $r^{in}$ where $\phi_{A}(r^{in})=\phi_{B}(r^{in})$. And the aggregation number of the B-block is defined as 
\begin{equation}
N^{agg}_{B} = \int_{0}^{r^{in}}4\pi r^{2}\phi_{B}(r)dr
\end{equation}

\section{Results}
\subsection{Cononsolvency Induced Micelle Behaviors at the Critical Point} 
Figure~\ref{fig:6_figure_1.png} show the polymer concentration and micelle aggregation number change upon the addition of cosolvents at the binodal boundary with $\chi_{BC}$ equal to $4$, $4.5$, $5$. The binodal boundary is defined as the point where the grand potential of the homogeneous system equals to the micellar system. The increase of the cosolvent quality, that is the decrease of the $\chi_{BC}$, will cause the critical polymer concentration curve shifting downwards, suggesting the expansion of the micelle phase. The strong B-block - cosolvent affinity interaction can also extend the cosolvent fraction range in which micelle can be formed as shown in figure~\ref{fig:6_figure_1.png} (a). The dependence of phase behavior on excess affinity strength is similar to homopolymer bulk system. The stronger the cosolvent excess affinity, the larger the inhomogeneous phase will become. \par

\Figure{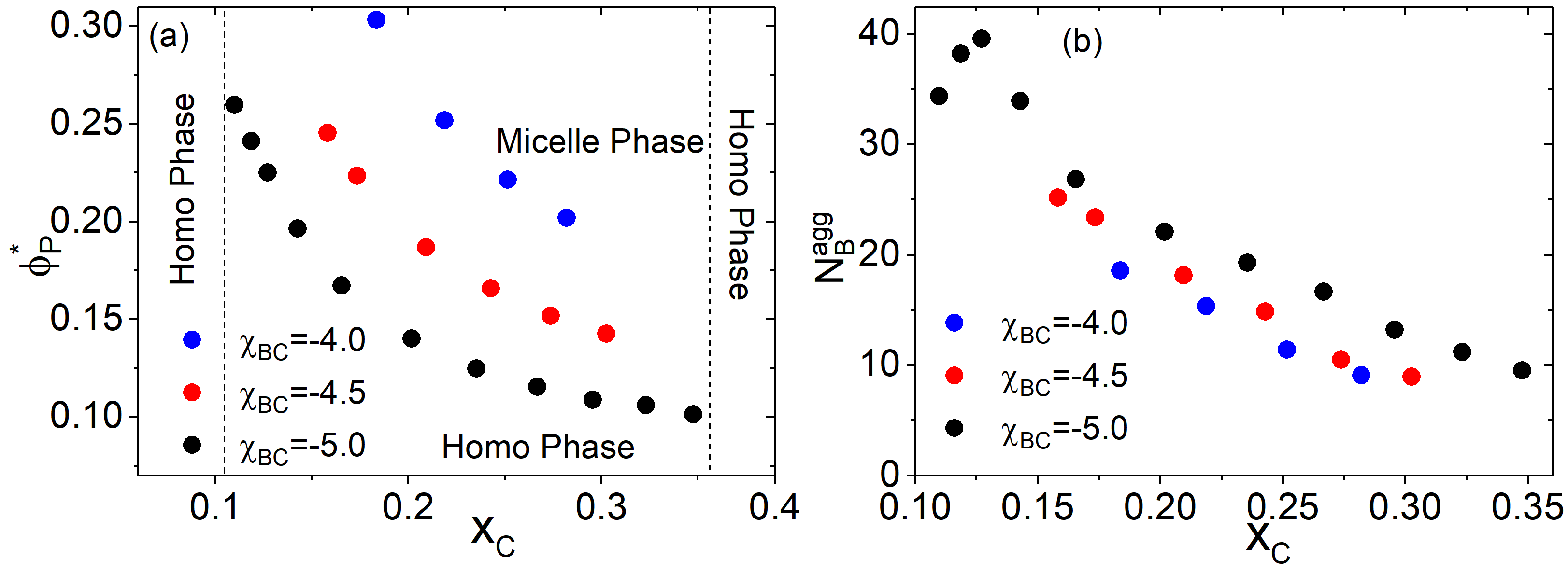}{0.98\linewidth}{(a) Critical micelle concentration as a function of cosolvent fraction at different B-block - cosolvent affinity strength. (b) Aggregation number of B-segments in the micelle core at different B-block - cosolvent affinity strength.}

The structural features of the micelles can be analyzed through aggregation number plots, which describes the number of B-block segments within the core and reflect micelle size variations due to cosolvent addition. In the $\chi_{BC}=-5$ system, the micelle size exhibits a nonmonotonic change with increasing cosolvent fraction, highlighting the cononsolvency effect, which initially intensifies and then diminishes — behavior similar to that observed in bulk polymer systems. Likewise, the most "collapsed" state induced by cononsolvency occurs at an intermediate cosolvent fraction, a trend also observed in homopolymer systems.
Another notable observation is that micelles form abruptly at a specific cosolvent fraction but gradually disappear with further cosolvent addition. This trend in chain conformation changes due to cononsolvency has been observed experimentally in both homopolymer and diblock copolymer systems \cite{ge2007stimuli, zhang2001reentrant}. It highlights a fundamental characteristic of cononsolvency: a sudden onset followed by a gradual decline. The aggregation number curve shifts slightly downwards with the increase of $\chi_{BC}$, indicating the lessening of cononsolvency. Additionally, At $\chi_{BC}=-4$ and $-4.5$, the micelle phase does not extend to lower cosolvent fractions, resulting in a monotonic micelle size change. The aggregation number and critical concentration plots suggest that A-B diblock copolymer systems share key cononsolvency features with homopolymer bulk systems.
\par

\subsection{Cononsolvency Induced Micelle Morphology Study} 
Next, we investigate density profiles in several systems with different cosolvent fraction and $\chi_{BC}$. All of system density profiles are measured at critical point, where the grand potential of homogeneous system equals to that of micellar system.\par
Figure~\ref{fig:6_figure_2.png} shows the morphology difference of conventional micelle induced by solvent selectivity with micelle structure induced by pure cononsolvency effect. In selectivity driven system, the micelle core composition predicted by SCFT calculation majorly consisted of B-block, which is very "dry" due to B-S repulsive interaction. And the observation is consistent with previous reports \cite{leermakers1995structure,patterson2013structural}. But the micelle driven by cononsolvency effect contains large amount of solvents and cosolvents in the core. The similar results have also been shown in experiments \cite{rao2007cononsolvency}. The B-block-cosolvent affinity interaction causes the enrichment of the cosolvent inside the core, and conversely, the solvents distribution is depleted in the core compared with its bulk density. The large amount of solvents/cosolvents inside the core may cause the looseness or softening of the micelle structure. Moreover, it may change its dynamics properties.  \par

\Figure{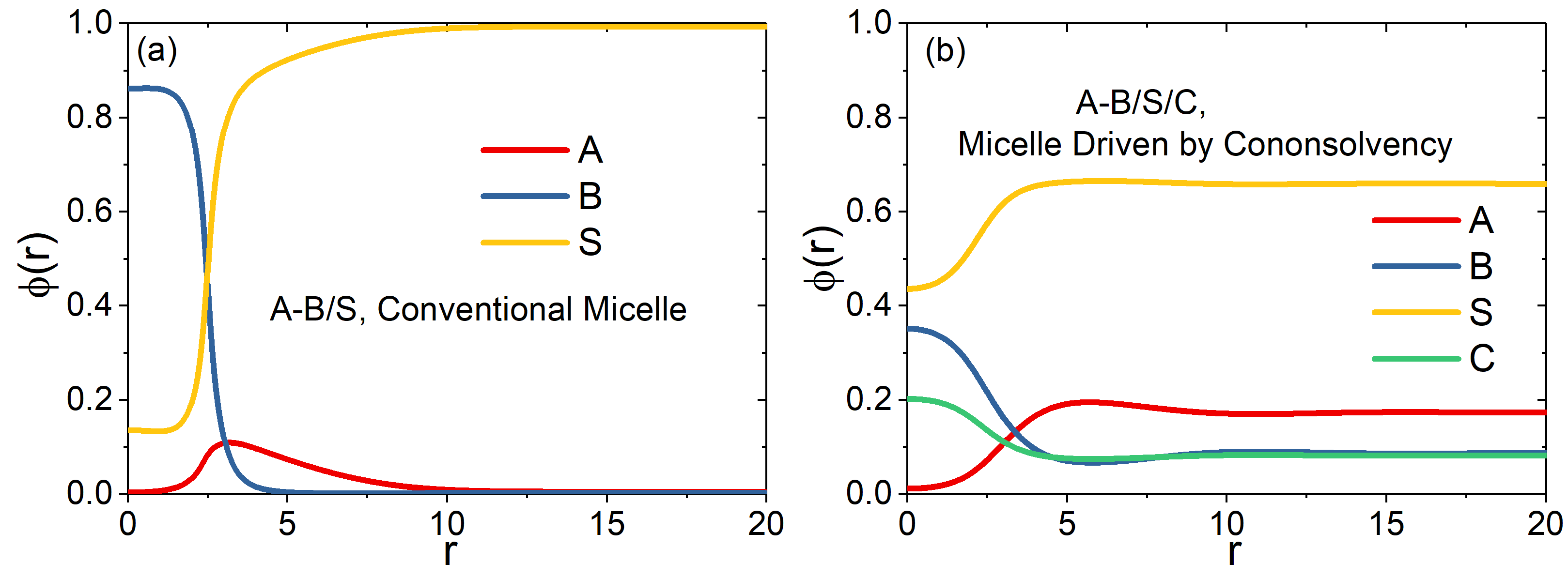}{0.98\linewidth}{(a) Density profile of conventional micelle (A-B/S system, $\chi_{BS}=1.5$), driven by B-block - solvents repulsion. (b) Density profile of micelle driven by cononsolvency effect (A-B/S/C system, $\chi_{BC}=-5, x_{C}=0.1$).}

The cosolvent quality effect on micelle structure can be characterized by plotting the density profile at the same cosolvent fraction with varied $\chi_{BC}$. Figure~\ref{fig:6_figure_3.png} shows the structure difference at $x_{C}=0.2$ with different $\chi_{BC}$ value. The similar micelle core composition distribution as figure~\ref{fig:6_figure_2.png} (b) can be observed. The distribution of solvents becomes depleted due to the preferential mixing between B-block segments and cosolvents. The micelle shell, which is consisted of the A-block, is quite extended because of the good solvent condition. With the increase of the B-block-cosolvent affinity strength, or in other words, the increase of the cosolvent quality, the promoted inhomogeneity can be deduced from the improved depletion of the solvents and enhanced enrichment of cosolvents inside the core. Correspondingly, the largest aggregation number of B-block segments can be observed in strongest B-C affinity interaction system. \par

\Figure{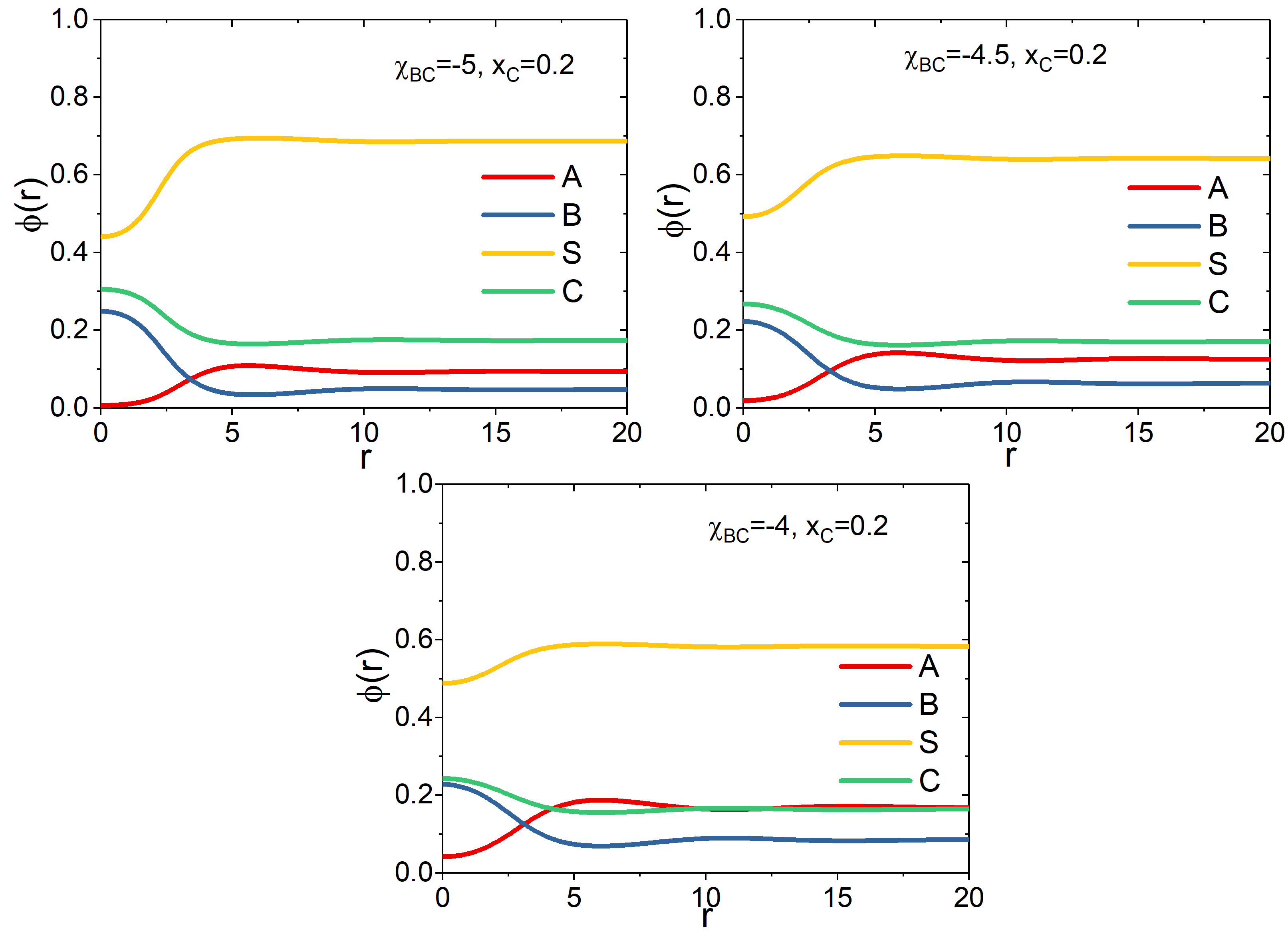}{0.98\linewidth}{Density profile of micelle system at cosolvent fraction equal to $0.2$ with different B-block - cosolvents attractive interactions ($\chi_{BC}=-5, \chi_{BC}=-4.5, \chi_{BC}=-4$).}

The density profile variation at the homogeneous-micellar phase transition point with $\chi_{BC}=-5$ upon the addition of cosolvents is presented in figure~\ref{fig:6_figure_4.png}. It can be found that the micelle with highly aggregated structure emerges as long as cosolvent fraction reaches a critical value. But the system becomes homogeneous gradually with the addition of the cosolvents. And the nonmonotonic variation trend indicated by aggregation number plot is not clear in density profile change, as the degree of increase of $N^{agg}_{B}$ is small. \par

\Figure{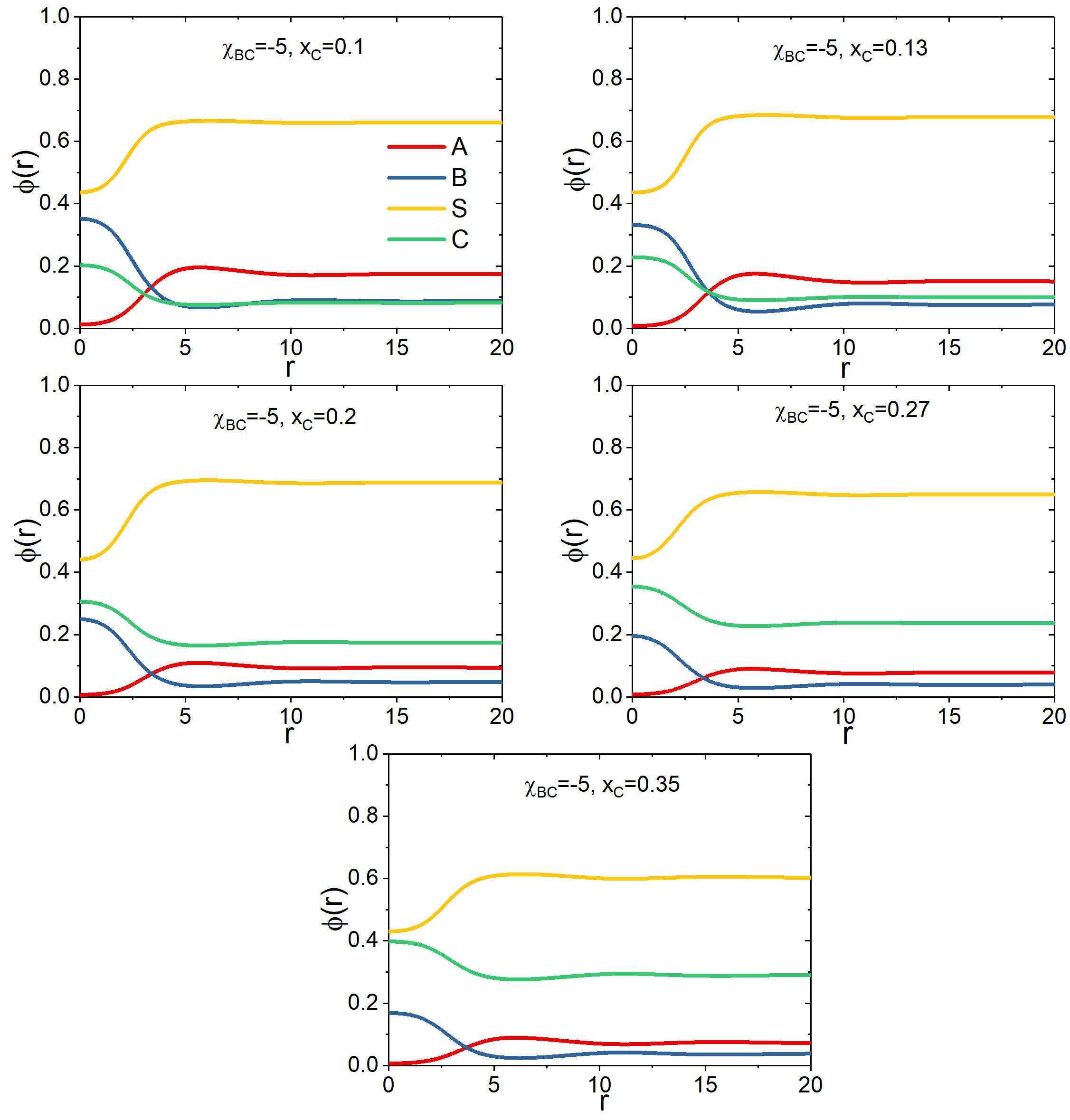}{0.98\linewidth}{Evolution of density profile for $\chi_{BC}=-5$ system with the increase of the cosolvent fraction.}

\subsection{The Driving Force for Cononsolvency Induced Micelle} 
It has been argued that the formation of micelle due to cononsolvency effect has the similar thermodynamics driving force as its amphiphilic block copolymer analogies, as they both try to reduce the interfacial tension \cite{michailova2018self}. But the SCFT free energy results suggests the totally different mechanism. The free energy difference is calculated by subtracting grand potential of homogeneous system from that of micellar system. X-Y contact area is obtained by dividing the X-Y energy with X-Y interaction strength, which is $\chi_{XY}$. Figure~\ref{fig:6_figure_5.png} (a) shows the total free energy difference (TFED) and B-segments - solvent contact area difference (CAD) for conventional selectivity solvent driven micelle. With the increase of the polymer concentration, the micelle system becomes more stable, showing much lower total free energy than homogeneous system. The decrease of the TFED accompanies with the decrease of the B-S CAD, indicating the reduction of the B-S interaction caused by polymer aggregation plays the major role to stabilize the micelle. And that is the general view for conventional micelle \cite{mai2012self,zhulina2012theory}. But in cononsolvency effect driven micelle, the micelle stabilization process is associated with the increase of the B-C CAD as it is shown in figure~\ref{fig:6_figure_5.png} (b), signifying that the system total free energy is minimized by the maximization of the B-block-cosolvent interaction, leading to the polymer aggregates. This is significantly different from the usual understanding on conventional micelle. \par

\Figure{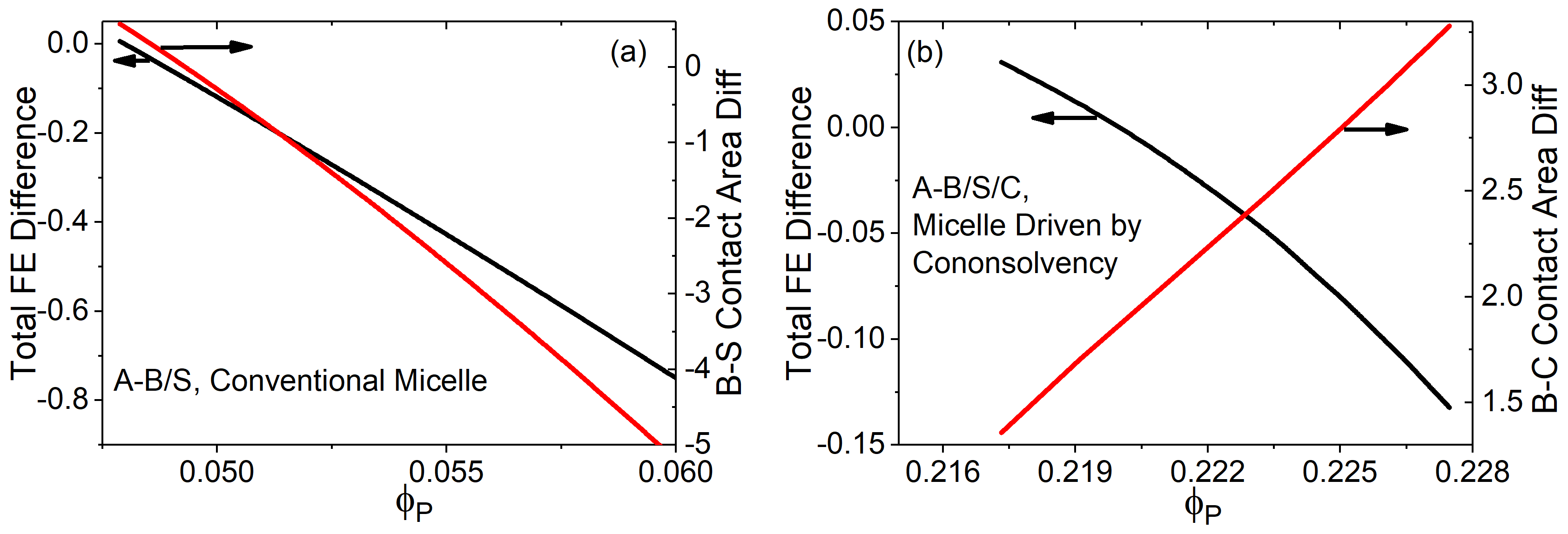}{0.98\linewidth}{(a) The free energy and B-block - Solvents contact area difference (micelle system - homogeneous system) of conventional micelle plotted against the polymer concentration ($\chi_{BS}=1.5$). (b) The free energy and B-block - Cosolvents contact area difference (micelle system - homogeneous system) of micelle driven by cononsolvency effect plotted against polymer concentration ($\chi_{BC}=-5, x_{C}=0.13$).}

\section{Discussion}
The structure of the micelle driven by cononsolvency effect is remarkably different from that of the solvent selectivity driven micelle. Hence, it must have an effect on some other properties, like chain exchange. 
The high free energy barrier can be observed at the position close to the micelle radius when the chain is pulled away from the micelle core \cite{seeger2021free}. The contribution to the free energy barrier arise from both A-B incompatibility and core-solvent repulsive interaction, as the solvents and shell-block are both enriched in the micelle shell and the magnitude of the barrier highly depends on the core-block - solvent $\chi$ value \cite{seeger2021free}. Hence, it can be deduced that the chain exchange process in cononsolvency driven micelle system will be much more encouraged than that of the conventional micelle system, because the A-B incompatibility becomes the major force to impair the chain exchange when the polymer chain drifts into the shell as solvents and cosolvents are both intrinsically good solvents for the polymer. And the related study about cononsolvency micelle chain transportation will be done in the future research. The investigation of the structure of the cononsolvency driven micelle can help investigate the molecules transportation process, and in further, it will help promote DHBCs' applications in nano-container, drug delivery, etc.
\par
The ordered micro-structure can originate from either repulsive interaction difference, like amphiphilic polymer, or affinity interaction difference, like phase transition induced by cononsolvency, which is investigated in this study \cite{mai2012self,schmidt2018double,bharadwaj2022cononsolvency}. It has been shown that the loose spherical aggregates can be formed by double hydrophilic block copolymer in neat solvents due to affinity difference \cite{casse2012solution,park2016micelles}. So, it is interesting to compare the self-assembly of DHBCs driven by cononsolvency effect (A-B/S/C system) and self-assembly of DHBCs in pure solvents system (A-B/S system), as both of them result from affinity difference, but in different pair interactions. In our SCFT calculation, the micro-structure cannot be observed in pure cosolvent system, even with strong B-C attractive interaction, like $\chi_{BC}=-5$. And experimental results show that unimer-micelle-unimer transition is observed due to the occurrence of cononsolvency effect, signifying that micelle is not formed in pure solvents or pure cosolvents \cite{rao2007cononsolvency}. So, there should be some significant difference between cononsolvency driven micelle and micelle formed by DHBCs in A-B/S system. \par
Based on experimental data, it strongly suggests that the core of the A-B/S system micelle is consisted of the block which is less hydrophilic, and the block with higher affinity to the solvents dwells in the shell \cite{casse2012solution,park2016micelles}. This is opposite to cononsolvency induced micelle, whose core is consist of block with stronger attractive interaction to cosolvents, and whose shell is made up of relatively less hydrophilic block. In this study, we show that the increase of the core-block - cosolvent (with better quality than solvents) contact area is responsible for the minimization of the total free energy of cononsolvency driven micelle. So, we can deduce that the driving force for the micellization of double hydrophilic block copolymer in pure solvents is still the reduction of the core-block - solvents contact area, like conventional selectivity driven micelle. If the mechanism of it is the same as cononsolvency driven micelle, the core-block will be the more hydrophilic block, but this is not what is observed in experiments. In general, both of these two types of micellization arise from the affinity difference of one block to the solvent (or cosolvent), but the driving force is totally different. 

\section{Conclusion}
The micellization can be induced by merely cononsolvency effect. The solvent/cosolvent fraction and their quality difference can have an effect on micelle phase behaviors and structures. The larger the quality difference is, the more profound micelle structure will be. The increase of the quality difference can also expand the micellar phase. The structure and the driving force of the cononsolvency induced micelle are compared with the conventional solvent selectivity driven micelle. The micelle core contains a large amount of solvent and cosolvent in cononsolvency induced micelle. Conversely, the core is very "dry" in conventional micelle. And the driving force for cononsolvency induced micelle is the maximization of the core-block - cosolvent(with better quality than solvents) contact area, which is totally different from conventional micelle. These differences may give rise to different chain exchange behaviors, which is discussed in the above section. The self-assembled structure of DHBCs in a pure solvent is compared with cononsolvency-driven micelles, revealing that, despite both being driven by affinity differences, their underlying driving forces are fundamentally different.

\newpage
\bibliographystyle{unsrt}
\bibliography{citation}

\begin{thebibliography}{10}

\bibitem{sistach2010thermoresponsive}
St{\'e}phanie Sistach, Mariana Beija, Virginie Rahal, Annie Br{\^u}let,
  Jean-Daniel Marty, Mathias Destarac, and Christophe Mingotaud.
\newblock Thermoresponsive amphiphilic diblock copolymers synthesized by
  madix/raft: properties in aqueous solutions and use for the preparation and
  stabilization of gold nanoparticles.
\newblock {\em Chemistry of Materials}, 22(12):3712--3724, 2010.

\bibitem{khimani2020self}
Mehul Khimani, Hiren Patel, Vijay Patel, Paresh Parekh, and Rohit~L Vekariya.
\newblock Self-assembly of stimuli-responsive block copolymers in aqueous
  solutions: An overview.
\newblock {\em Polymer Bulletin}, 77(11):5783--5810, 2020.

\bibitem{wei2009thermo}
Hua Wei, Si-Xue Cheng, Xian-Zheng Zhang, and Ren-Xi Zhuo.
\newblock Thermo-sensitive polymeric micelles based on poly
  (n-isopropylacrylamide) as drug carriers.
\newblock {\em Progress in Polymer Science}, 34(9):893--910, 2009.

\bibitem{ge2007stimuli}
Zhishen Ge, Dang Xie, Daoyong Chen, Xiaoze Jiang, Yanfeng Zhang, Hewen Liu, and
  Shiyong Liu.
\newblock Stimuli-responsive double hydrophilic block copolymer micelles with
  switchable catalytic activity.
\newblock {\em Macromolecules}, 40(10):3538--3546, 2007.

\bibitem{li2008self}
Guiying Li, SEN Song, LEI Guo, and Songmei Ma.
\newblock Self-assembly of thermo-and ph-responsive poly (acrylic acid)-b-poly
  (n-isopropylacrylamide) micelles for drug delivery.
\newblock {\em Journal of Polymer Science Part A: Polymer Chemistry},
  46(15):5028--5035, 2008.

\bibitem{chung1999thermo}
JE~Chung, M~Yokoyama, M~Yamato, T~Aoyagi, Y~Sakurai, and T~Okano.
\newblock Thermo-responsive drug delivery from polymeric micelles constructed
  using block copolymers of poly (n-isopropylacrylamide) and poly
  (butylmethacrylate).
\newblock {\em Journal of Controlled Release}, 62(1-2):115--127, 1999.

\bibitem{zhang2014polymer}
Jingli Zhang, Mingxi Zhang, Kangjian Tang, Francis Verpoort, and Taolei Sun.
\newblock Polymer-based stimuli-responsive recyclable catalytic systems for
  organic synthesis.
\newblock {\em Small}, 10(1):32--46, 2014.

\bibitem{schmidt2018double}
Bernhard~VKJ Schmidt.
\newblock Double hydrophilic block copolymer self-assembly in aqueous solution.
\newblock {\em Macromolecular Chemistry and Physics}, 219(7):1700494, 2018.

\bibitem{papadakis2019switch}
Christine~M Papadakis, Peter M{\"u}ller-Buschbaum, and Andr{\'e} Laschewsky.
\newblock Switch it inside-out:“schizophrenic” behavior of all
  thermoresponsive ucst--lcst diblock copolymers.
\newblock {\em Langmuir}, 35(30):9660--9676, 2019.

\bibitem{qian2019new}
Sijia Qian, Shenzhen Li, Weifeng Xiong, Habib Khan, Jing Huang, and Wangqing
  Zhang.
\newblock A new visible light and temperature responsive diblock copolymer.
\newblock {\em Polymer Chemistry}, 10(36):5001--5009, 2019.

\bibitem{liu2014dual}
Xueshibojie Liu, Dan Yu, Chunshun Jin, Xiaowei Song, Jinzhang Cheng, Xue Zhao,
  Xinmeng Qi, and Guangxin Zhang.
\newblock A dual responsive targeted drug delivery system based on smart
  polymer coated mesoporous silica for laryngeal carcinoma treatment.
\newblock {\em New Journal of Chemistry}, 38(10):4830--4836, 2014.

\bibitem{guragain2015multi}
Sudhina Guragain, Bishnu~Prasad Bastakoti, Victor Malgras, Kenichi Nakashima,
  and Yusuke Yamauchi.
\newblock Multi-stimuli-responsive polymeric materials.
\newblock {\em Chemistry--A European Journal}, 21(38):13164--13174, 2015.

\bibitem{rapoport2007physical}
Natalya Rapoport.
\newblock Physical stimuli-responsive polymeric micelles for anti-cancer drug
  delivery.
\newblock {\em Progress in Polymer Science}, 32(8-9):962--990, 2007.

\bibitem{wang2008investigation}
Huan Wang, Yingli An, Nan Huang, Rujiang Ma, and Linqi Shi.
\newblock Investigation of the cononsolvency effect on micellization behavior
  of polystyrene-b-poly (n-isopropylacrylamide).
\newblock {\em Journal of colloid and interface science}, 317(2):637--642,
  2008.

\bibitem{wolf1975measured}
BA~Wolf and G~Blaum.
\newblock Measured and calculated solubility of polymers in mixed solvents:
  Monotony and cosolvency.
\newblock {\em Journal of Polymer Science: Polymer Physics Edition},
  13(6):1115--1132, 1975.

\bibitem{wolf1978measured}
BA~Wolf and MM~Willms.
\newblock Measured and calculated solubility of polymers in mixed solvents:
  Co-nonsolvency.
\newblock {\em Die Makromolekulare Chemie: Macromolecular Chemistry and
  Physics}, 179(9):2265--2277, 1978.

\bibitem{rao2007cononsolvency}
Jingyi Rao, Jian Xu, Shizhong Luo, and Shiyong Liu.
\newblock Cononsolvency-induced micellization of pyrene end-labeled diblock
  copolymers of n-isopropylacrylamide and oligo (ethylene glycol) methyl ether
  methacrylate.
\newblock {\em Langmuir}, 23(23):11857--11865, 2007.

\bibitem{michailova2010nanoparticles}
V~Michailova, I~Berlinova, P~Iliev, L~Ivanov, S~Titeva, G~Momekov, and
  I~Dimitrov.
\newblock Nanoparticles formed from pnipam-g-peo copolymers in the presence of
  indomethacin.
\newblock {\em International journal of pharmaceutics}, 384(1-2):154--164,
  2010.

\bibitem{kyriakos2016quantifying}
Konstantinos Kyriakos, Martine Philipp, Che-Hung Lin, Margarita Dyakonova,
  Natalya Vishnevetskaya, Isabelle Grillo, Alessio Zaccone, Anna Miasnikova,
  Andr{\'e} Laschewsky, Peter M{\"u}ller-Buschbaum, et~al.
\newblock Quantifying the interactions in the aggregation of thermoresponsive
  polymers: the effect of cononsolvency.
\newblock {\em Macromolecular Rapid Communications}, 37(5):420--425, 2016.

\bibitem{ko2021co}
Chia-Hsin Ko, Cristiane Henschel, Geethu~P Meledam, Martin~A Schroer, Renjun
  Guo, Luka Gaetani, Peter M{\"u}ller-Buschbaum, Andr{\'e} Laschewsky, and
  Christine~M Papadakis.
\newblock Co-nonsolvency effect in solutions of poly (methyl
  methacrylate)-b-poly (n-isopropylacrylamide) diblock copolymers in
  water/methanol mixtures.
\newblock {\em Macromolecules}, 54(12):5825--5837, 2021.

\bibitem{dudowicz2015communication}
Jacek Dudowicz, Karl~F Freed, and Jack~F Douglas.
\newblock Communication: Cosolvency and cononsolvency explained in terms of a
  flory-huggins type theory.
\newblock {\em The Journal of chemical physics}, 143(13):131101, 2015.

\bibitem{cowie1987alcohol}
John~MG Cowie, Mahmood~A Mohsin, and Iain~J McEwen.
\newblock Alcohol-water cosolvent systems for poly (methyl methacrylate).
\newblock {\em Polymer}, 28(9):1569--1572, 1987.

\bibitem{wang2012conformational}
Fei Wang, Yi~Shi, Shuangjiang Luo, Yongming Chen, and Jiang Zhao.
\newblock Conformational transition of poly (n-isopropylacrylamide) single
  chains in its cononsolvency process: a study by fluorescence correlation
  spectroscopy and scaling analysis.
\newblock {\em Macromolecules}, 45(22):9196--9204, 2012.

\bibitem{zhang2001reentrant}
Guangzhao Zhang and Chi Wu.
\newblock Reentrant coil-to-globule-to-coil transition of a single linear
  homopolymer chain in a water/methanol mixture.
\newblock {\em Physical review letters}, 86(5):822, 2001.

\bibitem{knop2010poly}
Katrin Knop, Richard Hoogenboom, Dagmar Fischer, and Ulrich~S Schubert.
\newblock Poly (ethylene glycol) in drug delivery: pros and cons as well as
  potential alternatives.
\newblock {\em Angewandte chemie international edition}, 49(36):6288--6308,
  2010.

\bibitem{casse2012solution}
Olivier Casse, Andriy Shkilnyy, J{\"u}rgen Linders, Christian Mayer, Daniel
  H{\"a}ussinger, Antje V{\"o}lkel, Andreas~F Th{\"u}nemann, Rumiana Dimova,
  Helmut C{\"o}lfen, Wolfgang Meier, et~al.
\newblock Solution behavior of double-hydrophilic block copolymers in dilute
  aqueous solution.
\newblock {\em Macromolecules}, 45(11):4772--4777, 2012.

\bibitem{ke2009association}
Fuyou Ke, Xiulei Mo, Runmiao Yang, Yanmei Wang, and Dehai Liang.
\newblock Association of block copolymer in nonselective solvent.
\newblock {\em Macromolecules}, 42(14):5339--5344, 2009.

\bibitem{rudolph2013synthesis}
Tobias Rudolph, Sarah Crotty, Moritz Von~der L{\"u}he, David Pretzel, Ulrich~S
  Schubert, and Felix~H Schacher.
\newblock Synthesis and solution properties of double hydrophilic poly
  (ethylene oxide)-block-poly (2-ethyl-2-oxazoline)(peo-b-petox) star block
  copolymers.
\newblock {\em Polymers}, 5(3):1081--1101, 2013.

\bibitem{willersinn2017aqueous}
Jochen Willersinn and Bernhard~VKJ Schmidt.
\newblock Aqueous self-assembly of pullulan-b-poly (2-ethyl-2-oxazoline) double
  hydrophilic block copolymers.
\newblock {\em Journal of Polymer Science Part A: Polymer Chemistry},
  55(22):3757--3766, 2017.

\bibitem{park2016micelles}
H~Park, S~Walta, RR~Rosencrantz, A~K{\"o}rner, Christoph Schulte, L~Elling,
  Walter Richtering, and Alexander B{\"o}ker.
\newblock Micelles from self-assembled double-hydrophilic
  phema-glycopolymer-diblock copolymers as multivalent scaffolds for lectin
  binding.
\newblock {\em Polymer Chemistry}, 7(4):878--886, 2016.

\bibitem{willersinn2017self}
Jochen Willersinn and Bernhard~VKJ Schmidt.
\newblock Self-assembly of double hydrophilic poly (2-ethyl-2-oxazoline)-b-poly
  (n-vinylpyrrolidone) block copolymers in aqueous solution.
\newblock {\em Polymers}, 9(7):293, 2017.

\bibitem{michailova2018self}
Victoria~I Michailova, Denitsa~B Momekova, Hristiana~A Velichkova, Evgeni~H
  Ivanov, Rumiana~K Kotsilkova, Daniela~B Karashanova, Elena~D Mileva, Ivaylo~V
  Dimitrov, and Stanislav~M Rangelov.
\newblock Self-assembly of a thermally responsive double-hydrophilic copolymer
  in ethanol--water mixtures: the effect of preferential adsorption and
  co-nonsolvency.
\newblock {\em The Journal of Physical Chemistry B}, 122(22):6072--6078, 2018.

\bibitem{wu2022synthesis}
Ruonan Wu, Yanru Chen, Jing Zhou, and Yebang Tan.
\newblock Synthesis, characterization and application of dual thermo-and
  solvent-responsive double-hydrophilic diblock copolymers of
  n-acryloylmorpholine and n-isopropylacrylamide.
\newblock {\em Journal of Molecular Liquids}, 357:119053, 2022.

\bibitem{kyriakos2014cononsolvency}
Konstantinos Kyriakos, Martine Philipp, Joseph Adelsberger, Sebastian Jaksch,
  Anatoly~V Berezkin, Dersy~M Lugo, Walter Richtering, Isabelle Grillo, Anna
  Miasnikova, Andr{\'e} Laschewsky, et~al.
\newblock Cononsolvency of water/methanol mixtures for pnipam and ps-b-pnipam:
  pathway of aggregate formation investigated using time-resolved sans.
\newblock {\em Macromolecules}, 47(19):6867--6879, 2014.

\bibitem{zhou2016pfs}
Hang Zhou, Yijie Lu, Meng Zhang, Gerald Guerin, Ian Manners, and Mitchell~A
  Winnik.
\newblock Pfs-b-pnipam: A first step toward polymeric nanofibrillar hydrogels
  based on uniform fiber-like micelles.
\newblock {\em Macromolecules}, 49(11):4265--4276, 2016.

\bibitem{bharadwaj2022cononsolvency}
Swaminath Bharadwaj, Bart-Jan Niebuur, Katja Nothdurft, Walter Richtering, Nico
  van~der Vegt, and Christine~M Papadakis.
\newblock Cononsolvency of thermoresponsive polymers: where we are now and
  where we are going.
\newblock {\em Soft Matter}, 2022.

\bibitem{zhang2020unified}
Xiangyu Zhang, Jing Zong, and Dong Meng.
\newblock A unified understanding of the cononsolvency of polymers in binary
  solvent mixtures.
\newblock {\em Soft Matter}, 16(33):7789--7796, 2020.

\bibitem{zhang2024general}
Xiangyu Zhang, Jing Zong, and Dong Meng.
\newblock General condition for polymer cononsolvency in binary mixed solvents.
\newblock {\em Macromolecules}, 57(17):8632--8642, 2024.

\bibitem{leermakers1995structure}
FAM Leermakers, CM~Wijmans, and GJ~Fleer.
\newblock On the structure of polymeric micelles: Self-consistent-field theory
  and universal properties for volume fraction profiles.
\newblock {\em Macromolecules}, 28(9):3434--3443, 1995.

\bibitem{patterson2013structural}
Joseph~P Patterson, Elizabeth~G Kelley, Ryan~P Murphy, Adam~O Moughton,
  Mathew~P Robin, Annhelen Lu, Olivier Colombani, Christophe Chassenieux, David
  Cheung, Millicent~O Sullivan, et~al.
\newblock Structural characterization of amphiphilic homopolymer micelles using
  light scattering, sans, and cryo-tem.
\newblock {\em Macromolecules}, 46(15):6319--6325, 2013.

\bibitem{mai2012self}
Yiyong Mai and Adi Eisenberg.
\newblock Self-assembly of block copolymers.
\newblock {\em Chemical Society Reviews}, 41(18):5969--5985, 2012.

\bibitem{zhulina2012theory}
EB~Zhulina and OV~Borisov.
\newblock Theory of block polymer micelles: recent advances and current
  challenges.
\newblock {\em Macromolecules}, 45(11):4429--4440, 2012.

\bibitem{seeger2021free}
Sarah~C Seeger, Kevin~D Dorfman, and Timothy~P Lodge.
\newblock Free energy trajectory for escape of a single chain from a diblock
  copolymer micelle.
\newblock {\em ACS Macro Letters}, 10(12):1570--1575, 2021.

\end{thebibliography}

\end{document}